\documentclass[prb,twocolumn,showpacs,superscriptaddress]{revtex4}

\usepackage{graphicx}
\usepackage{amsmath}
\usepackage{amssymb}
\usepackage{epsfig}

\renewcommand{\v}[1]{{\bf #1}}

\def\eqa{\begin{eqnarray}}
\def\eea{\end{eqnarray}}
\newcommand{\eq}{\begin{equation}}
\newcommand{\ee}{\end{equation}}
\newcommand{\nn}{\nonumber\\}

\newcommand{\<}{\langle}
\renewcommand{\>}{\rangle}

\newcommand{\al}{\alpha}
\newcommand{\bt}{\beta}

\newcommand{\Del}{\Delta}

\newcommand{\ga}{\gamma}

\newcommand{\si}{\sigma}

\usepackage{color}

\begin{document}

\title{The direction of the $d$-vector in a nematic triplet superconductor}

\author{Lin Yang}
\affiliation{National Laboratory of Solid State Microstructures $\&$ School of Physics, Nanjing
	University, Nanjing, 210093, China}

\author{Qiang-Hua Wang}
\email{qhwang@nju.edu.cn}
\affiliation{National Laboratory of Solid State Microstructures $\&$ School of Physics, Nanjing
	University, Nanjing, 210093, China}
\affiliation{Collaborative Innovation Center of Advanced Microstructures, Nanjing 210093, China}

\begin{abstract}
We investigate the states of triplet pairing in a candidate nematic superconductor versus typical material parameters, using the mean field theory for two- and three-dimensional tight-binding models with local triplet pairing in the $E_u$ representation. In the two-dimensional model, the system favors the fully gapped chiral state for weaker warping or lower filling level, while a nodal and nematic $\Del_{4x}$ state is favorable for stronger warping or higher filling, with the $d$-vector aligned along the principle axis. In the presence of lattice distortion, relative elongation along one of the principle axes, $\v a$, tends to rotate the nematic $d$-vector orthogonal to $\v a$, resulting in the nematic $\Del_{4y}$ state at sufficient elongation. Three-dimensionality is seen to suppress the chiral state in favor of the nematic ones. Our results may explain the variety in the probed direction of the $d$-vector in existing experiments.
\end{abstract}

%

\maketitle


\section{Introduction}

Time-reversal-invariant (TRI) topological superconductors (TSC) attract sustained interests due to the interesting Majorana zero modes in vortices and itinerant Majorana fermions on the boundary.\cite{Schnyder,Qi1,RRoy} It has been established that the key requirement for TSC in inversion symmetric systems is odd-parity pairing.\cite{Fu1,Sato}
Although considerable efforts are made both theoretically and experimentally,\cite{Qi2,Nakosai,Scheurer,Wang,Hosur}, definite evidence of a TRI-TSC is yet to be found. A best candidate to date is $\mathrm{Cu_xBi_2Se_3}$,\cite{Hor} which is made by intercalation of Cu into $\mathrm{Bi_2Se_3}$. The parent compound is a topological insulator (TI). While the topology of the normal state is not necessary for TSC,\cite{Fu1} the strong spin-orbital coupling (SOC) in such a material makes TSC more likely. The maximum transition temperature $T_c$ observed is 3.8K. Early specific heat measurement \cite{Kriener} seems to indicate a full pairing gap. The upper critical field exceeds the Pauli limit,\cite{Bay} suggesting triplet pairing. Assuming odd parity triplet pairing, a candidate pairing function is $\phi(\v k)=\tau_2 \si_3 i\si_2$,\cite{Fu1} dubbed $\Del_2$ pairing. Henceforth, $\tau_i$ ($\si_i$) is the Pauli matrix in the orbital (spin) basis. The two orbitals are derived from the $p_z$ orbitals of Bi-atoms in a quintable layer of Bi$_2$Se$_3$. The $\Del_2$ pairing is a nondegenerate representation of $D_{3d}$. The $d$-vector is along $z$ and the SC state is fully gapped. The zero-bias conductance peak observed in the point-contact spectroscopy indicates the existence of unusual in-gap surface states.\cite{Sasaki} However, the scanning-tunneling-microscopy (STM) measurement reveals a full gap with no sign of in-gap surface states.\cite{Levy} With the spectroscopic uncertainty in mind,
recent nuclear magnetic resonance (NMR) experiment \cite{Matono} makes a breakthrough in this field. The observed Knight-shift $K$ develops a two-fold oscillation as a function of the angle of the in-plane applied field $H$, with strongest (or no) suppression of $K$ below $T_c$ for $H$ along (or orthogonal to) one of the Se-Se bonds, the principle axes henceforth. This could be understood if the $d$-vector of the triplet aligns along a principle axis. Fu realized that the inplane NMR nematicity suggests the pairing function must belong to a doublet representation.\cite{Fu3} For local pairing, the desired pairing function is obvious, $\phi(\v k)=\tau_2 \si_{1,2} i\si_2$, dubbed $(\Del_{4x},\Del_{4y})$ pairing, with the understanding that the $d$-vector is along and orthogonal to the principle axis, respectively. Note that $\Del_{4x}$ leads to nodal SC gap, protected by the $D_{3d}$ group, while $\Del_{4y}$ could become fully gapped in the presence of warping effect in the normal state band structure.\cite{Fu3}  While the nematicity is observed by various probes in Cu$_x$Bi$_2$Se$_3$, the identified direction of the $d$-vector varies.\cite{Matono,Yonezawa,Pan,Smylie,Asaba,haihu,donglai} Theoretically,\cite{baowc} direct visualization of the $d$-vector is possible by quasi-particle interference (QPI) \cite{qpi} and STM: the leading peak momentum in QPI at sub-gap energies should be along the $d$-vector, and the STM profile of the vortex at low energies should be elongated also along the $d$-vector. The agreement between the results in the momentum space (from QPI) and real space (from vortex profile) is a stringent constraint for the nematic triplet. 

In real samples, there may be lattice distortions, \cite{distortion} which could pin the direction of the $d$-vector. On the other hand, the extent of warping, the filling level, the thickness of the sample, as well as the strength of inter-layer hybridization, may vary from sample to sample, and their roles for the $d$-vector are to be unravelled.  
Here we study how the nematic pairing, and the direction of the $d$-vector in particular, depends on the above typical material parameters, in order to understand the variety in the probed $d$-vector direction in existing experiments. We use a mean field theory (MFT) based on a tight-binding model, assuming local triplet pairing in the $E_u$ representation. 

Our main results are as follows. In the two-dimensional (2D) model, the system favors the fully gapped chiral state for weaker warping or lower filling level, while a nodal and nematic $\Del_{4x}$ state is favorable for stronger warping or higher filling. In the presence of lattice distortion, relative elongation along one of the principle axes, tends to rotate the nematic $d$-vector in favor of the nematic $\Del_{4y}$ state. In the 3D model, increasing inter-layer hybridization suppresses the chiral state in favor of the nematic ones. 

The rest of the paper is organized as follows. The model is described in Sec.\ref{sec:model}, the effect of warping in the 2D model is described in Sec.\ref{sec:2d}, the effect of lattice distortion in Sec.\ref{sec:distortion}, and the effect of inter-layer hybridization in Sec.\ref{sec:3d}. Finally, Sec.\ref{sec:summary} is a summary of this work.

\section{Model and methods}\label{sec:model}

\begin{figure}
	\includegraphics[width=0.9\columnwidth]{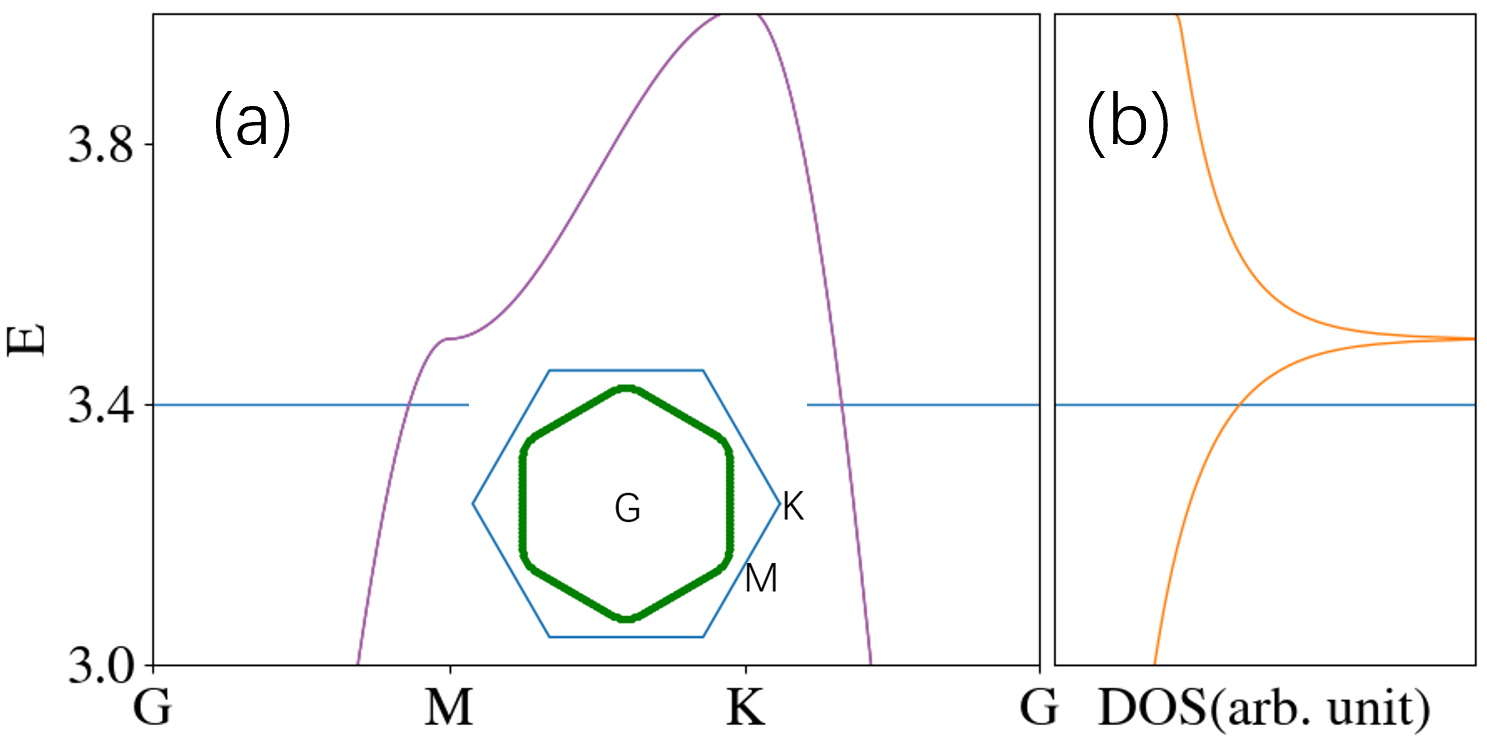}
	\caption{The band structure of the 2D model for $\mu=3.4$, $m=-0.5$ and $t_w=0.1$. (a) The band dispersion along high-symmetry cuts and near the Fermi level (the horizontal line). The inset shows the Fermi surface (green line) in the Brillouine zone (outer hexagon), with high symmetry momenta $G$, $K$ and $M$ indicated. (b) The density of states.}\label{fig:band}
\end{figure}

According to the first-principles calculations,\cite{Zhang} the conduction and valence bands of $\mathrm{Bi_2Se_3}$ are superpositions of Se $p_z$ orbitals on the top and bottom layers of the unit cell, each of which mixed with it's neighboring Bi $p_z$ orbital. The angle-resolved photoemission spectroscopy (ARPES) experiment\cite{Wray} shows that the band structure of $\mathrm{Cu_xBi_2Se_3}$ is quite similar to $\mathrm{Bi_2Se_3}$. A two-orbital continuum model \cite{Fu3} has been proposed to describe the low energy physics of $\mathrm{Cu_xBi_2Se_3}$. Here we use a similar minimal model defined on the layered triangular lattice, which is equivalent to the continuum model at low energy. The free part of the Hamiltonian can be written as, in the momentum space, $H_0=\sum_{\v k}\psi_{\v{k}}^\dag h_{\v{k}} \psi_{\v{k}}$, where $\psi_{\v k}$ is a four-component spinor describing two orbitals and two spins, with $\psi_{\v{k}}^\dag=(c_{\v k1\uparrow}^\dag,c_{\v k2\uparrow}^\dag,c_{\v k1\downarrow}^\dag,c_{\v k2\downarrow}^\dag)$, and
\eqa h_{\v k}=&&\sum_i \al_i (\v d_i\times \pmb{\sigma})_z \tau_3 \sin k_i + t_w\sum_i \al_i \sigma_3\tau_3 f_i \sin k_i \nn
&&+[m+\sum_i \al_i (1-\cos k_i) + t_z(1 - \cos k_z)]\tau_1 \nn
&&+ t_z \sin k_z \tau_2-\mu. \label{eq:hk}\eea
Here the summation is over the three inplane translation vectors $\v d_1=(1,0)$, $\v d_2=(1/2,\sqrt{3}/2)$ and $\v d_3=(-1/2,\sqrt{3}/2)$, $k_i=\v k\cdot \v d_i$; $t_w$ is the warping parameter allowed in a $D_{3d}$ system, induced here by the form factor $f_{1,2,3}=(1,-1,1)$; $\mu$ is the chemical potential; and $m$ controls the topology of the normal state band structure. The 3D dispersion is controlled by the interlayer hybridization $t_z$, and in the 2D limit we set $t_z=0$. Finally, the coefficient $\al_i=1$ unless specified explicitly otherwise (for the distorted lattice).
Throughout this work we use arbitrary units for qualitative purposes, and we fix $m=-0.5$ for concreteness. 

As an example, we show the band structure in the 2D limit in Fig.\ref{fig:band}(a) for $\mu=3.4$ and $t_w=0.1$, which is consistent with the ARPES measurement.\cite{Wray} The corresponding density of states (DOS) is shown in Fig.\ref{fig:band}(b). The DOS has a Van Hove singularity near the Fermi level, and this makes the system sensitive to perturbations that we will be addressing. 

The NMR nematicity in the SC state strongly implies triplet pairing in a doublet $E_u$ channel. While the underlying pairing mechanism is not yet clear, for our purpose it is sufficient and reasonable to start with an effective MF Hamiltonian with attractive pairing interaction in the $E_u$ channel,
\eqa H_{\rm MF}=H_0+\frac{1}{2}\sum_{\v k}\left(\Psi_\v k^\dag \tau_2\v V\cdot\vec\si i\si_2\Psi_{-\v k}^{\dag,t}+{\rm h.c.}\right),\eea
where $\Psi_\v k^\dag = (\psi_\v k^\dag, \psi_{-\v k}^T)$ is the Nambu spinor, $t$ means transpose, and $\v V=(\Del_{4x},\Del_{4y})$ is the $d$-vector order parameter determined self-consistently by
\eqa \Del_{4x,4y}=\frac{V_s}N \sum_{\v k} \<\Psi_{-\v k}^t (-i\si_2) \si_{1,2} \tau_2 \Psi_\v k\>, \eea
where $V_s <0$ is the pairing interaction and $N$ is the number of lattice sites. The local triplet is made possible by the antisymmetric pairing between the two orbitals (or orbital singlet). The chiral pairing is defined by $\v V\times \v V^*\neq 0$, while the nematic pairing is signaled by $\v V\times \v V^*= 0$. The Bogoliubov-de Gennes (BdG) quasiparticles, with energy dispersion $E_\v k$, can be obtained by diagonalizing the single-particle part of $H_{\rm MF}$. In the following we define the BdG gap at a polar angle as $|E_\v k|_{\rm min}$ for $\v k$ along the corresponding radial direction in the inplane Brillouin zone. Usually this is just $E_\v k$ on the normal state Fermi surface (FS), but in the present model the unusual pairing may cause $|E_\v k|_{\rm min}$ to deviate from the FS. 

\section{Two dimensional limit}\label{sec:2d}

\begin{figure}
	\includegraphics[width=0.9\columnwidth]{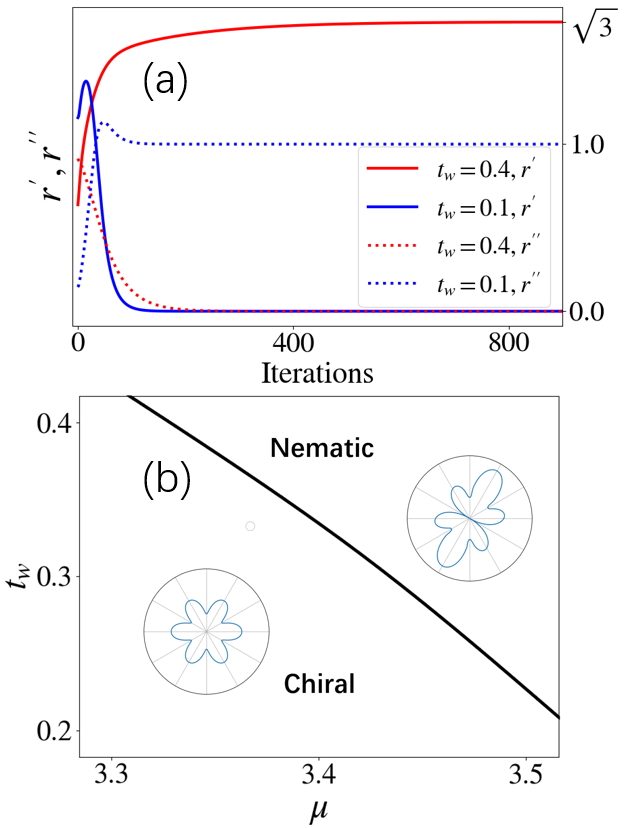}
	\caption{(a) The evolution of the real part ($r'$, solid lines) and imaginary part ($r''$, dashed lines) of the ratio $r=\Delta_{4y}/\Delta_{4x}$ versus MF iteration steps for $t_w=0.1$ (blue) and $t_w=0.4$ (red). Here $\mu=3.4$, $V_s=-1.5$, and $T=10^{-5}$. (b) The phase diagram obtained by MF calculations in the $(\mu, t_w)$ space with $V_s=-1.5$. The insets illustrate the characteristic BdG gap versus the polar angle in the respective phases.}\label{fig:phase} 
\end{figure}

First, we consider the 2D limit, $t_z=0$. We fix $(\mu, V_s)=(3.4, -1.5)$ and perform the MF calculation at the temperature $T=10^{-5}$, the zero temperature limit. The solid (dotted) lines in Fig.\ref{fig:phase}(a) show the MF evolution of the real (imaginary) part of the ratio $r=\Del_{4y}/\Del_{4x}$, for $t_w=0.1$ (blue) and $t_w=0.4$ (red). Clearly, a chiral state with $\v V\propto (1, i)/\sqrt{2}$ is obtained for $t_w=0.1$. An equivalent state is $\v V\propto (1,-i)/\sqrt{2}$. On the other hand, a nematic state with $\v V\propto (1,\sqrt{3})/2$ is obtained for $t_w=0.4$. There is a discrete six-fold degeneracy in the nematic states, with
\eqa \Delta_{4y}/\Del_{4x}=\tan(n\pi/3),\ \ n=1,2,...,6. \label{eq:6fold}\eea
The value of $n$ in the converged MF solution depends on the initial condition. 
Both the chiral state and the six-fold degenerate $\Del_{4x}$ state can be captured by a phenomenological Landau free energy in the form
\eqa f=&&\alpha~ \v V^*\cdot \v V+\beta (\v V^*\cdot\v V)^2 +\beta'|\v V\times \v V^*|^2 \nn
+&& \gamma [(\Del_{4x}^*+i\Del_{4y}^*)^3(\Del_{4x}+i\Del_{4y})^3+{\rm c.c.}]+\cdots.\eea
With $\al<0$ and $\bt>0$ as usual, the chiral phase is stable for $\bt'<0$ and $\ga= 0$,\cite{chiral} while the nematic phase is stable for $\bt'>0$ and $\gamma <0$. Note the 6-th order $\ga$-term is required to reproduce the discrete six-fold degeneracy in the nematic state. Interestingly, a similar six-fold degeneracy of helical triplet pairing was found theoretically in doped BiH, but a 12-th order term is needed instead.\cite{yanglin}  

By systematic MF calculations with $V_s=-1.5$ and in the zero temperature limit, we obtain a phase diagram, Fig.\ref{fig:phase}(b), for the pairing state in the $(\mu, t_w)$ parameter space. Note a higher value of $\mu$ corresponds to a higher electron filling. We see that the chiral state is favorable for lower warping or filling, and the nematic state is favored otherwise. We stress that the nematic state is of $\Del_{4x}$-type. The $\Del_{4y}$-type pairing is not obtained in the MF calculations here. The characteristic BdG gap veresus the polar angle is shown as insets in the respective phase, which is nodeless/nodal in the chiral/nematic phase.

\section{Lattice distortion}\label{sec:distortion}

\begin{figure}
	\includegraphics[width=0.9\columnwidth]{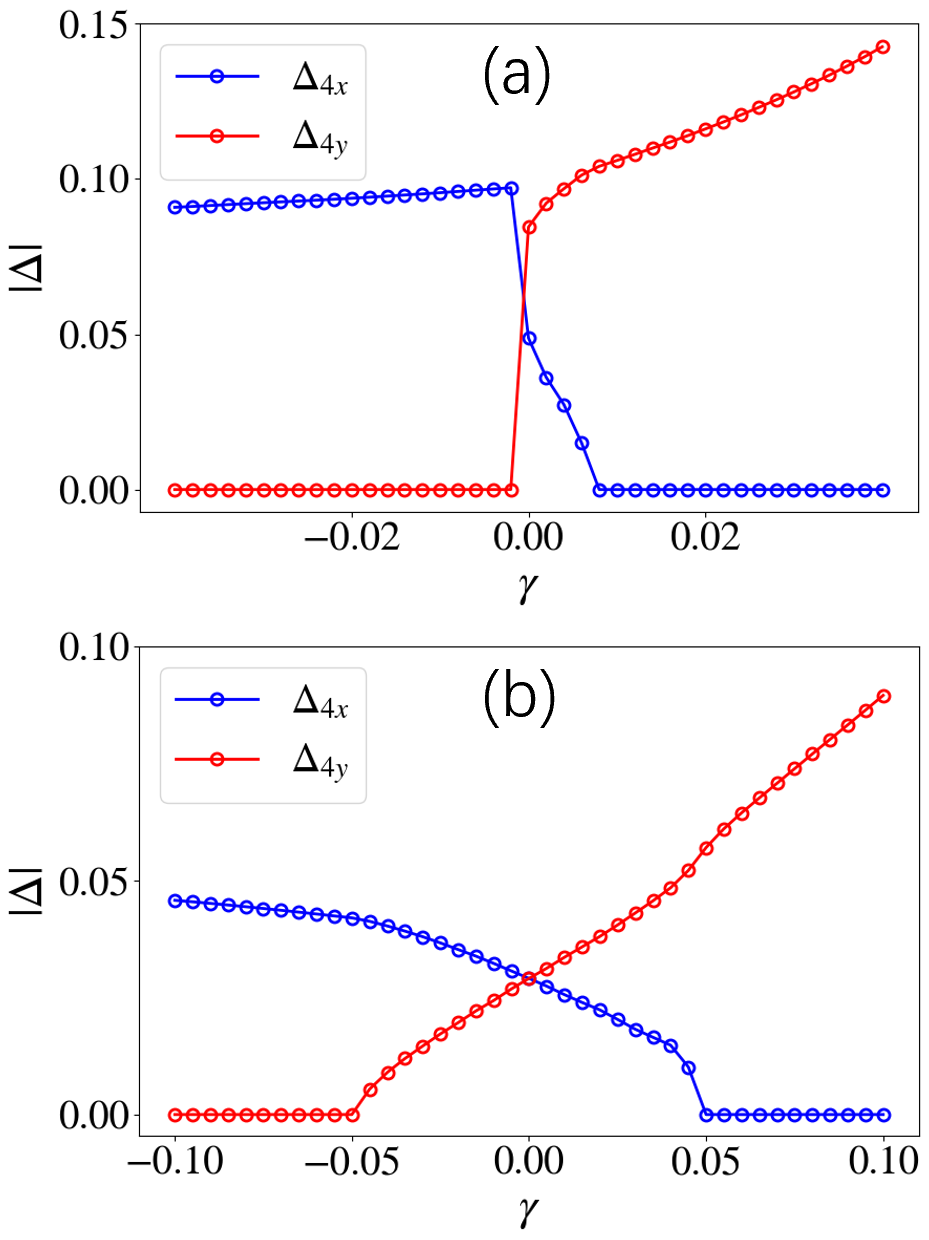}
	\caption{The amplitude of $\Del_{4x}$ and $\Del_{4y}$ versus the distortion parameter $\ga$, for (a) $(\mu,t_w)=(3.4,0.4)$, and (b) $(\mu,t_w)=(3.4,0.1)$.}\label{fig:distortion}
\end{figure}

In the previous section, we find the $\Del_{4y}$ pairing is not stabilized in the ideal 2D lattice model. However, this type of pairing state was reported in experiments. We then ask how it could be realized by perturbations away from the ideal limit. One possible perturbation is the lattice distortion, which is indeed observed in Sr$_x$Bi$_2$Se$_3$ by X-ray diffraction.\cite{distortion} By symmetry, the effect of the lattice distortion on the hopping coefficients $\al_{1,2,3}$ defined in Eq.\ref{eq:hk} can be decomposed into two symmetry channels, $\Del\al_i\propto x_i^2 - y_i^2$ and $\Del\al_i\propto 2 x_i y_i$. Here $(x_i,y_i)=\v d_i$ is the two components of the first-neighbor bond $\v d_i$. We concentrate on the $d_{x^2-y^2}$ channel. (The other channel can be discussed similarly.) Although all of $\al_{1,2,3}$ would be changed, we can perform rescaling so that $\al_{2}=\al_3=1$, and $\al_1=1-\gamma$, to represent the qualitative effect of the strain. A positive $\ga$ corresponds to relative elongation of $\v d_1$ versus $\v d_{2,3}$, and vice versa. (The rescaling may also change $m$ and $\mu$, but for a small distortion such changes could be ignored in the leading order approximation.)

Fig.\ref{fig:distortion}(a) shows the MF solution versus $\ga$ in the parent nematic case with $(\mu, t_w)=(3.4,0.4)$. The $\Delta_{4y}$ component increases rapidly for positive $\ga$, and the $\Del_{4x}$ component decreases to zero already for $\ga\geq 0.008$. We checked that $\Del_{4y}/\Del_{4x}$ remains real, so the pairing is still nematic. We see that the relative elongation along $x$ can rotate the $d$-vector to the orthogonal direction, in favor of the $\Del_{4y}$-pairing. For small positive $\ga$, the $d$-vector interpolates between $x$- and $y$-directions. Such an intermediate state may have been observed in Ref.\onlinecite{donglai}. Note for a negative $\ga$, or relative compression of the lattice along $x$, the $\Del_{4x}$ state is the only stable one. This is reasonable since the system favors $\Delta_{4x}$ already in the parent undistorted lattice for $(\mu, t_w)=(3.4,0.4)$. 

In the parent chiral case with $(\mu, t_w)=(3.4,0.1)$, the effect of $\ga$ on the $\Del_{4x,4y}$ components of $\v V$ is shown in Fig.\ref{fig:distortion}(b). The two components coexist within $|\ga|\leq 0.05$, and we checked that $\Del_{4y}/\Del_{4x}$ remains imaginary. So in this regime the pairing is still chiral. However, only $\Del_{4y}$ ($\Del_{4x}$) is left for $\ga>0.05$ ($\ga<-0.05$), hence the pairing becomes nematic. Although the critical value of $\ga$ for the transition between chiral and nematic states is relatively large, we again see that the relative elongation of the $x$-bonds in the 2D lattice tends to favor $\Del_{4y}$ pairing, and vice versa. By symmetry, the conclusion can be generalized to the relative elongation of any one of the principle axes of the triangle lattice. 

\section{Dimension crossover}\label{sec:3d}

\begin{figure}
	\includegraphics[width=0.9\columnwidth]{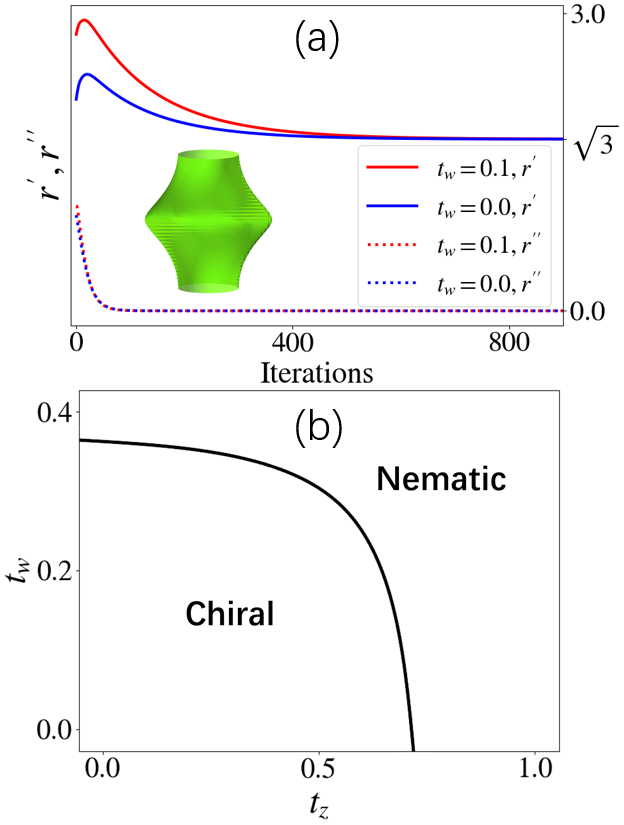}
	\caption{MF results for the 3D model with $\mu=3.4$ and $V_s=-2.5$. (a) The evolution of the real part ($r'$, solid lines) and imaginary part ($r''$, dashed lines) of the ratio $r=\Delta_{4y}/\Delta_{4x}$ versus MF iteration steps for $t_w=0$ (blue) and $t_w=0.1$ (red). Here $t_z=1.0$. The inset shows the FS. (b) The phase diagram in the $(t_z, t_w)$ space.}\label{fig:phase3d}
\end{figure}

It is known that at low doping, Cu$_x$Bi$_2$Se$_3$ has an ellipsoidal Fermi pocket centered at $\Gamma$, while at high doping the FS is open and cylinder-like along $k_z$, as observed in the photoemission\cite{fermi1} and de Haas-van Alphen measurements.\cite{fermi2, fermi3} In our tight-binding model, the shape of the 3D FS is controlled by the inter-layer hybridization $t_z$. For small $t_z$, the FS is open and cylinder-like. For $t_z>2$, the FS becomes closed around the $\Gamma$ point. We should remark that in the case of an open cylinder-like FS, both types of nematic pairing are topologically trivial, irrespectively of whether they are nodal or nodeless, since the FS encloses two time-reversal invariant momenta. With this in mind, we continue to investigate the pairing state in the 3D limit. We model the 3D lattice by $N_z=10$ layers of triangle lattices and periodic boundary condition along all translation directions. 

As an example of the 3D model, we consider $\mu = 3.4$, $t_z = 1.0$ and $V_s=-2.5$. (In the 3D model, the DOS is lower. To get a sizable $T_c$, a stronger pairing strength is used.) Fig.\ref{fig:phase3d}(a) shows the FS in the inset, and the evolution of the MF solutions in the main panel, for the real (solid line) and imaginary (dotted line) parts of the ratio $r=\Del_{4y}/\Del_{4x}$. We find in both cases of $t_w=0$ (blue lines) and $t_w=0.1$ (red lines), the solution converges to $\v V=(\Del_{4x},\Del_{4y})\propto (1,\sqrt{3})/2$, namely the $\Del_{4x}$ state up to a $C_6$ rotation. 

By systematic calculations, we obtain the phase diagram in the $(t_z, t_w)$ parameter space, see Fig.\ref{fig:phase3d}(b).  The chiral state is limited to smaller $t_z$ or $t_w$, while the nematic $\Del_{4x}$ phase prevails for larger $t_z$ or $t_w$. In reality, the hopping along $z$ is between quintuple layers, and should be weak. In this case, the result at large $t_w$, {\em e.g.}, $t_w=0.4$, is qualitatively similar to the 2D case discussed in Sec.\ref{sec:2d}. 

We note that the phenomenological Landau theory analysis in Ref.\onlinecite{chiral2} reaches a similar conclusion that reduced 3D dispersion would favor the chiral state, however, it does not include the effect of warping, hence is unable to capture the nematic phase in the 2D limit. On the other hand, we have further checked the effect of lattice distortion in the 3D model. The qualitative effect is the same as in Fig.\ref{fig:distortion}(a), namely, a relative elongation along $x$ would favor the $\Del_{4y}$ pairing.  

\section{Summary}\label{sec:summary}

To conclude, we studied the pairing states in a model for doped Bi$_2$Se$_3$ superconductor with triplet pairing in the $E_u$ representation. In the 2D model, the fully gapped chiral state is favored if the warping parameter is small, while the nodal nematic triplet with the $d$-vector $\v V=(\Del_{4x},0)$ is favored otherwise. Under lattice distortion, a relative elongation along $x$ would favor a $d$-vector $\v V=(0,\Del_{4y})$. In the 3D model, the chiral state disappears for large interlayer hopping, in favor of the nematic $\Del_{4x}$ state, and the effect of lattice distortion is similar to the case of the 2D model. Taking the above material parameters into account, our results may provide a possible cause of the variety in the probed $d$-vector direction in existing experiments.

\acknowledgements{The project was supported by the National Key Research and Development Program of China (under grant No. 2016YFA0300401) and the National Natural Science Foundation of China (under Grant No.11574134).}

\end{document}